\documentstyle[BoxedEPSF]{article}
\begin{document}
{{\bf Spontaneous Magnetisation in a Quantum Wire}}

\bigskip
\noindent
V.V'yurkov  and A.Vetrov\\
Institute of Physics and Technology, Russian
Academy of Sciences,\\
Nakhimovsky prosp. 34, Moscow, 117218, Russia\\
Phone: (095)3324918, Fax: (095)1293141
E-mail: vyurkov@ftian.oivta.ru

\bigskip
\it{
An existence of predominant symmetrical spin configuration (spin polarised
 phase) and "diluted" density of states (pseudo-gap) in a layer under the Fermi
 level in a quantum wire is predicted. The condition of cross-over
 from non-polarised phase to polarised one was derived. The transition occurs for 
sufficiently low electron density in a wire and is accompanied by an acute 
decrease of electron density of states near the Fermi level.It may result in a
corresponding decrease of conductance. A similar effect may exist in a two-dimensional
 electron gas.}
\bigskip

\rm
A lot of papers were issued on the theoretical and experimental 
investigation of low-dimensional systems: quantum wells, wires 
and dots, and many remarkable phenomena were revealed.  Along 
with a scientific attractiveness of such structures their especial properties
 might be applied in future microelectronic and optoelectronic devices. 
Here we concentrate on one of strong correlation effects in quantum wires 
(QWRs). These very effects can drastically change the properties of electron 
system with respect to non-interacting electron gas.

One of the most interesting problems arisen recently was connected with so
 called "0.7 structure". Firstly this miraculous structure was seen in the 
experiment with a quantum wire structure in 1996 \cite{1}. There was registered 
a pronounced additional step of a quantum wire conductance quantisation at 
the level 0.7 of a conductance quantum $G_{0} = 2e^2/{h}$.  A bit later 
an apparent
 deviation from a conductance quantum was also observed in the most perfect
 for today long QWRs \cite{2}. They were fabricated by cleaved overgrowth of
 GaAs/AlGaAs quantum well heterostructures. Even a decrease down to 50\% 
was revealed for $20\mu$ long wire. Worth mentioning that multiple quantization
steps visible in the experiment were subject to equal decrease. Later the
 fractional conductance quantum steps were seen in the experiments of the 
Copenhagen group [3-5] and even in metallic nanostructures \cite{6}.

Recently a novel observation of the 0.7 structure was made in quantum 
wires manufactured by split-gate technology \cite{7}.The authors saw 
again an additional step of quantization at the level 
$0.7G_{0}$. The fact that quite different structures revealed 
the same effect pointed out to its fundamental origin. 

For a while there was no adequate explanation consistent with all 
available experimental data. Attempts to apply the spin and spinless 
Luttinger liquid theories seemed the most appropriate to unravel 
the problem. Indeed, these theories gave the corrections to QWR 
conductance caused by Coulomb interaction between electrons \cite{8}:
\begin{equation}
G = G_{0} {(1 + V(0)/ \pi v_{F} )}^{-1/2}
\end{equation}		
where  $V(0)$  represents the Fourier transform $V(q)$ of the real 
space interaction potential between electrons for the transfer 
momentum q equal to zero, $v_{F}$  is the Fermi velocity. However, 
these corrections have monotonic dependence on the Fermi energy 
that contradicts with abrupt transition to common integer steps of 
conductance quantum with rising a gate voltage observed in the 
experiment \cite{7} and flat plateaux observed in \cite{2}. 

When a disorder was involved \cite{9} this also gave rise to an obvious 
decrease of the conductance but dependent on the electron density 
in the wire. The calculations fulfilled for realistic QWR wall 
roughness also revealed a strong dependence of scattering rate on 
the Fermi energy and subband number \cite{10}. Thus such a scattering 
can not be at all a feasible reason of conductance deviation because 
it do not accord with observed flat plateaus (within $5\%$) and 
conductance steps of equal height . 

Moreover, the latest experiments discovered an obvious connection 
of the "0.7 structure" with spin polarization of electrons in a 
QWR\cite{7}. There was seen a smooth transition of the "0.7 structure" for zero 
magnetic field to the "0.5 structure" when a magnetic field was 
going up \cite{3}. This evidence crucially sustains 
the hypothesis of spontaneous spin polarization
 of electrons in a QWR firstly put forward in Ref.\cite{11} prior to 
the experiment \cite{7}. Here we also argue that the spontaneous spin 
polarisation owes to exchange interaction between electrons in a QWR. 

Unfortunately, few papers were so far issued on the topic. A 
phenomenological explanation of a spontaneous magnetisation of a
 quasi-one dimensional conducting channel embedded in a Wigner crystal
 was presented in Ref.\cite{12}. In another Ref.\cite{13} the spontaneous 
polarisation (magnetisation) was merely postulated but its influence 
on a conductance was thoroughly discussed.

We start the consideration from a two-electron problem. The exchange
 energy is assumed to be small compared with kinetic energy. Therefore,
 we employ the Hartree-Fock (HF) approach to describe exchange interaction.
 
As for Coulomb interaction, it can be put into account in 
an audible self-consistent way. However, a realistic Coulomb 
potential in a QWR should be used \cite{14}. Surely,
 the electrostatic potential induced by internal electrons 
in a QWR can even blocade the wire conductance. But we adhere to 
the experimental conditions when the wire was quite penetrable 
for electrons and electrostatic potential can not influence on 
the linear response to infinitesimal bias applied to the wire.

    According to the HF approach the two-particle 
wave function looks like 

\begin{equation}
\Psi(x_1,x_2)={1\over\sqrt{2}}(\psi_1(x_1)\psi_2(x_2)\pm{\psi_1(x_2)\psi_2(x_1)})
\end{equation}
where $\psi_1,\psi_2$ are one-particle wave functions normalised per
 one electron in a wire ; $x_1, x_2$  are the co-ordinates along the wire. 
The upper sign corresponds to a singlet state (total spin equals S=0) 
and the lower sign corresponds to a triplet (S=1). 

   The HF equations for $\psi_1$ and $\psi_2$ are as follows

$$
\left(-{{h^2\over{2m}}}{d^2\over{dx^2}}+{{{e^2}\over{\kappa}}\int_{0}^L
{{{\vert\psi_2(x')\vert}^2}\over\vert{x-x'}\vert}dx'}\right){\psi_1(x)
\pm{{ {e^2}\over{\kappa}}\int_{0}^L{{\psi^*_2(x')\psi_1(x')}\over\vert{x-x'}\vert}dx'}}
\psi_2(x)=E_1\psi_1(x)
$$
\begin{equation}
\end{equation}
$$
\left(-{{h^2\over{2m}}}{d^2\over{dx^2}}+{{{e^2}\over{\kappa}}\int_{0}^L
{{{\vert\psi_1(x')\vert}^2}\over\vert{x-x'}\vert}dx'}\right){\psi_2(x)
\pm{{ {e^2}\over{\kappa}}\int_{0}^L{{\psi^*_1(x')\psi_2(x')}\over\vert{x-x'}\vert}dx'}}
\psi_1(x)=E_2\psi_2(x)
$$

Here  $\kappa$ is a permittivity, $L$  is a wire length. The integrals with constant sign
 + correspond to Coulomb interaction. Other integrals describe exchange
 interaction and their sign is determined by spin configuration. 

   The Coulomb interaction in the equations (3) is considered as perturbation.
 The unperturbed electron wave function in a one mode quantum wire is 

\begin{equation}
\psi(k,x)={1\over\sqrt{L}}e^{ikx}
\end{equation}

   Suppose that two electrons move across QWR in the same direction 
(left or right moving fields) with sufficiently small longitudinal momentum 
discrepancy  $h\Delta k$ so that 
\begin{equation}
        	h\Delta k<h/\lambda 
\end{equation}

where $\lambda$ is an effective screening length ( $\lambda < L$) and 
$L$ is a wire length. According to equ.s (3) these electrons possess 
exchange energy almost as great as Coulomb one 

\begin{equation}
		(e^2/ \kappa L)ln( \lambda /d) 
\end{equation}
here $d$  is a QWR diameter.
 
In our calculations with equ.s (3) the Coulomb potential $V(x)=1/ 
\kappa x$ was cut off for distances x smaller than the wire diameter d and larger than 
effective screening length $\lambda$. 

For greater momentum mismatch than that given by the inequality (5) 
the exchange integrals involve fast oscillating 
functions and tend to zero. A sign of exchange energy depends on the 
spin configuration. If electrons have an antisymmetric spin configuration 
(total spin equals unity) then their space wave function is symmetric and 
the sign of exchange energy is positive, i.e. the same as that of Coulomb energy.
Otherwise, when a total spin equals zero, the exchange energy is negative and 
reduces total energy of electron system. For the sake of simplicity in further 
calculations we suppose the exchange energy to be equal to the Coulomb one
(6) when the condition (5) is true. Otherwise, it is supposed to equal zero.

This model of exchange interaction was used to solve many-electron problem. 
It was found out that due to exchange interaction the ground 
state ($T=0$) corresponding to the minimum of the total energy 
(including kinetic one) can be that of predominant symmetrical 
spin configuration for electrons near the Fermi level, i.e. 
spin polarized.  The condition of the cross-over 
from conventional unpolarized state to polarized one is as follows
\begin{equation}
\lambda_F ln(\lambda/d)>a_B
\end{equation}
where $\lambda_F=4a$ is the Fermi electron wave length, $a^{-1}$ is 
an electron density in a wire and $a_B=h^2\kappa/me^2$ is a Bohr
 radius. 

To some extent, the above relation reminds the condition of 
Wigner crystallisation in a quantum wire deduced in Ref.\cite{15}. 
However, the condition (7) is valid for greater electron density 
in a QWR if only $\lambda > a$. In other words, the spin polarisation 
precedes the Wigner crystallisation.

Once the condition (7) is met the polarized phase arises in the energy 
interval under the Fermi level
\begin{equation}
\delta \varepsilon = (e^{2} / \kappa \lambda) ln(\lambda /d)		
\end{equation}

The magnitude of $\delta \varepsilon$ equals the exchange energy 
per one Fermi electron. Worth mentioning that it does not depend 
upon Fermi energy of electrons in any subband of a QWR. This is 
in a good qualitative agreement with the experimental evidence 
for conductance corrections to be insensitive to the Fermi 
energy \cite{2}. The spin configuration associated with the polarised
state is sketched in Fig.

In polarised state the exchange energy $\varepsilon_{ex}(k)$ for electrons adjacent
 to the Fermi level linearly depends upon momentum 

\begin{equation}
\varepsilon_{ex}(k)=4(e^2/\kappa)(k-k_F+\Delta{k})ln(\lambda /d)
\end{equation}

where $k_F=2\pi/\lambda_F$ is the Fermi wave vector.
The total energy equals $\varepsilon=\varepsilon_{ex}(k)+h^2{k^2}/{2m}$
(the kinetic energy is added here). 

The relation (9) originates from a dependence of an electron exchange 
energy upon spin configuration of near-by electrons in k-space (Fig.1). 
Here we put attention to the important point of our consideration.
Although the resulting correction to the energy (consequently  
corrections to the Fermi velocity $v_F$) given by exp.(9) may be quite small,
 a derivative $d\varepsilon_{ex}/dk$ connected with density of 
states $\rho(\varepsilon)\sim(d\varepsilon/dk)^{-1}$ may be large. 

From exp. (9) we deduce the relative decrease of electron density of
 states in the energy interval $\delta\varepsilon$ under the Fermi level as 

\begin{equation}
{\Delta\rho}/ \rho \sim( \pi a_B k_F)^{-1}ln(\lambda /d).
\end{equation}

According to inequality (7) the relative decrease of density of states (10)
 cannot be less than 0.25 in a spin polarised phase. However, as the parameter
 in the right hand side of the exp.(10) is not small a non-perturbation approach should be developed to get a
 precise number.

The spontaneous magnetisation can affect various phenomena in which electrons
 at the Fermi level are involved. In particular, the "diluted" density of 
states (or pseudo-gap) may result in corresponding decrease of conductance. 
Although the main goal of the present paper is to make clear the origin of a 
spontaneous magnetisation we briefly discuss its influence upon a conductance. 
The ballistic current through the wire biased by a voltage V can be calculated 
in the conventional way [17]

\begin{equation}
I=e{\int\limits^{\varepsilon_F+eV/2}_{\varepsilon_F-eV/2}}v(\varepsilon)\rho(\varepsilon)
d(\varepsilon)\approx{e^2v_F\rho(\varepsilon_F)V}
\end{equation}

Thus we obtain the same relative correction to the conductance as that given by formula
 (10) for density of states. Much more elaborate calculation of the conductance was given
 in Ref. \cite{13}. However, the magnetisation was postulated there. It should be outlined 
that our approach gives no corrections of the kind supplied by the exp. (1) to a wire 
conductance in depolarised state. It concords at the best with numerous experiments where
 integer plateaux of conductance were observed in controversy to the Luttinger liquid 
theory predictions. 

Our estimations show that the condition of crossover (7) to polarised
phase was valid even for the top Fermi energies attained in the 
experiment [2] (unlike to that in [7]). We accepted for evaluations
a screening length $\lambda$ as a distance from the QWR to the nearest gate 
electrode and the wire diameter d consistent with subband spacing 
(20 meV) pointed out in [2]. Then we gained $\delta\varepsilon_F$ exceeding kT 
(for T about 1K).  When the temperature T was rising the polarised
 phase was smeared and the conductance quantum restated.  
This explains the abnormal temperature dependence of the QWR 
conductance seen in the experiment. When the bias V exceeds $\delta \varepsilon/e $
"undiluted" electrons were involved in the conductance and a 
conductance quantum restates too.

To be consistent with over-all experimental data a wire length 
should be introduced in the theory. The experiment \cite{2} revealed
 a quite weak dependence of the conductance deviation on the wire 
length, at least, sub-linear one. The conductance deviation 
was only doubled while the wire length varied from $1\mu$ to $20\mu$.
 To possibilities look like plausible. The first one is that
 in the experimental structure the wire diameter diminishes as
the wire lengthens. An indirect insinuation to this very 
dependence was that less negative gate voltage pinched off 
a longer wire. The second possibility is an interaction of a 
wire with leads which partially destroy the polarised phase 
in the pre-contact region. Although the leads were already 
modelled in [18-19] this consideration looks quite deficient 
yet and further attempts are required. Otherwise, the presence 
of leads would be an eternal artifice to fit the theory to
 the experiment.

It should be noted that a similar phenomenon seems quite 
possible to exist in a two-dimensional electron gas. 

In conclusion, an existence of predominant symmetrical spin 
configuration (spin polarized phase) and "diluted" density 
of states under the Fermi level in the quantum wire is 
predicted.  The reduction of quantum wire conductance is in 
agreement with recent experimental data.The condition of 
cross-over from non-polarised phase to polarised one was 
derived. The transition occurs for sufficiently low electron
 density in a wire and is accompanied by an acute decrease 
of electron density of states near the Fermi level 
(pseudo-gap) resulting in a corresponding decrease of
conductance. The effect crucially depends upon screening.\\

{\bf Acknowledgments}\\
The work was supported through the program "Physics of solid 
state nanostructures" (grant N 97-1077) and program 
"Prospective technologies and devices of micro- and 
nanoelectronics" (grant N 02.04.4.1.32.E.35) of the Russian
 Ministry of Science, and also through the Russian Basic
 Research Foundation (grant N 000100397).

\eject
\HideDisplacementBoxes
\BoxedEPSF{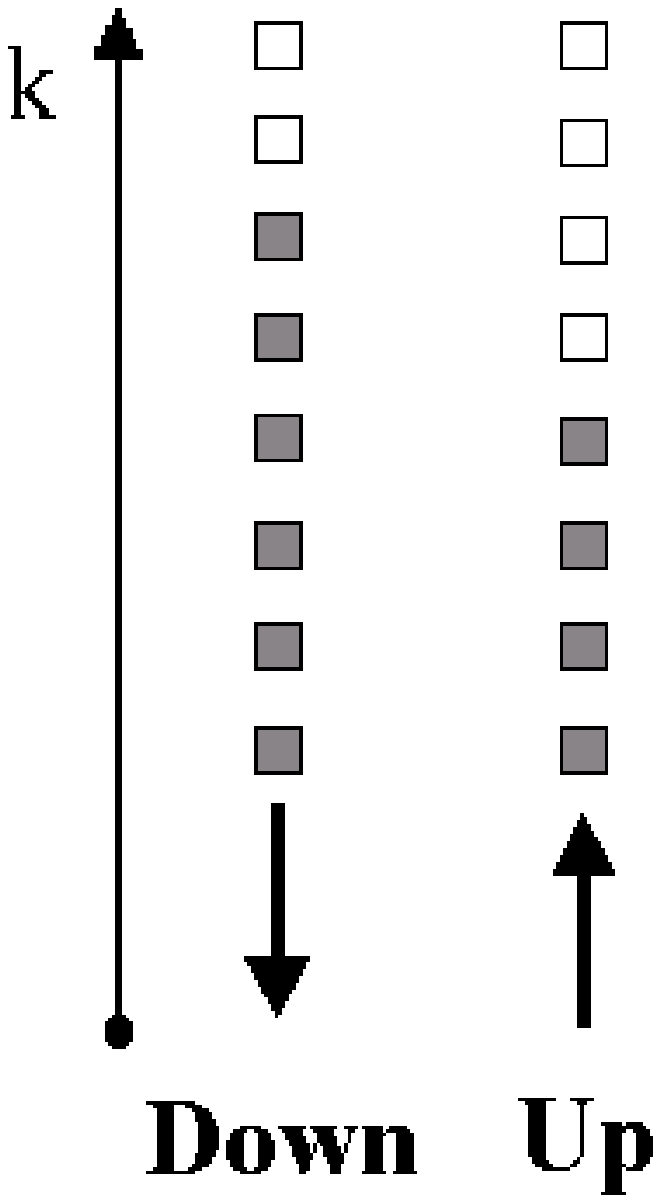}

FIGURE CAPTION.

Electrons with different orientation of spin (up and down) fill the longitudinal 
momentum k-space states near the Fermi level. The latter coincides with upper 
electron state in both stacks. Empty squares correspond to unoccupied states. 
The same situation exists for electrons moving in the opposite direction ($k<0$).

\end{document}